\RequirePackage{fix-cm}

\documentclass{svjour3}                     

\smartqed  
\usepackage{graphicx}
\usepackage{multirow}
\journalname{Theoretical Chemistry Accounts}

\usepackage{color}
\usepackage[version=3]{mhchem} 
\usepackage{comment}
\usepackage{geometry}
\begin{document}

\title{Infrared spectra of neutral polycyclic aromatic hydrocarbons by machine learning
}


\author{Ga\'etan LAURENS*\and
        Malalatiana RABARY \and
        Julien LAM \and
        Daniel PEL\'AEZ \and
        Abdul-Rahman ALLOUCHE*  
}


\institute{G. Laurens\at
            Institut Lumi\`ere Mati\`ere, UMR5306 Universit\'e Lyon 1-CNRS, Universit\'e de Lyon, 69622 Villeurbanne Cedex, France;\\
              \email{gaetan.laurens@univ-lyon1.fr}\\           
           \and
           M. Rabary\at
            Institut Lumi\`ere Mati\`ere, UMR5306 Universit\'e Lyon 1-CNRS, Universit\'e de Lyon, 69622 Villeurbanne Cedex, France
        \and
           J. Lam\at
            Center for Nonlinear Phenomena and Complex Systems, Code Postal 231, Université Libre de Bruxelles, Boulevard du Triomphe, 1050 Brussels, Belgium\\
            \and
           D. Pel\'aez\at
            Institut des Sciences Mol\'eculaires d'Orsay (ISMO) - UMR 8214. B\^at. 520, Universit\'e Paris-Saclay, 91405 Orsay Cedex
            \and
        A.R. Allouche\at
            Institut Lumi\`ere Mati\`ere, UMR5306 Universit\'e Lyon 1-CNRS, Universit\'e de Lyon, 69622 Villeurbanne Cedex, France\\
              \email{allouchear@univ-lyon1.fr} \\
}

\date{Received: date / Accepted: date}

\maketitle

\begin{abstract}
The Interest in polycyclic aromatic hydrocarbons (PAHs) spans numerous fields and infrared spectroscopy is usually the method of choice to disentangle their molecular structure. In order to compute vibrational frequencies, numerous theoretical studies employ either quantum calculation methods, or empirical potentials, but it remains difficult to combine the accuracy of the first approach with the computational cost of the second. In this work, we employed Machine Learning techniques to develop a potential energy surface and a dipole mapping based on an artificial neural network (ANN) architecture. Altogether, while trained on only 11 small PAH molecules, the obtained ANNs are able to retrieve the infrared spectra of those small molecules, but more importantly of 8 large PAHs different from the training set, thus demonstrating the transferability of our approach.


\end{abstract}

\section{Introduction}
\label{intro}
Polycyclic aromatic hydrocarbons (PAHs) is a family of highly-stable organic molecules presenting two or more fused aromatic rings. This stability implies their ubiquity in very different environments.
On the one hand, they are widespread in the interstellar medium (up to 15~\% of the total carbon)~\cite{Dwek1997Feb,Tielens2008Aug,Micelotta2010Feb}.
This fact, together with their spectroscopical features, motivated in the early eighties the so-called PAH-hypothesis by means of which PAH would be responsible for the Aromatic Infrared Bands~\cite{Candian2018}. 
On Earth, PAHs are also very abundant and are generated in incomplete combustion processes and are found in flames~\cite{Dobbins2006Mar,Lafleur1996Mar,Oktem2005Sep,Adamson2018Dec} and soot~\cite{Calcote1981Jan,Homann1985Jan,Frenklach1994,Frenklach2002May,Haynes1981Jan}.
The latter are complex clusters mainly composed of PAHs and related species which have been suggested as major contributors to the greenhouse effect~\cite{ramanathan2008global}.
In addition to all this, PAHs have attracted major attention owing to their environmental impact~\cite{holloway2010} and their implication in health issues~\cite{Downward2014Dec,Abdel-Shafy2016Mar,Duran2016Aug} including lung cancer~\cite{Downward2014Dec}. 
Consequently, major efforts have been devoted to the design of novel methods for the elimination of PAHs molecules~\cite{Kumari2018Apr,Jeelani2017Aug,Kalmykova2014Jun}.
In the meantime, the role of PAH dimerization in the soot nucleation constitutes a topic of major interest~\cite{Mercier2019Apr,Kholghy2018Apr,Faccinetto2020Aug,Kholghy2019Jan}.

Owing to their wide interest, numerous theoretical studies has been dedicated to molecular simulations of PAHs molecules~\cite {Yuan2019Feb,Chen2019Nov,Bauschlicher2018Feb,Michoulier2018Mar,Mao2017Sep}. From a modeling point of view, two types of approaches have been employed so far: quantum chemistry simulations, ranging from PAH cluster formation to spectroscopical studies~\cite{Michoulier2018Mar}, and empirical potentials based on ReaxFF interactions~\cite{Chen2019Nov,Yuan2019Feb,Mao2017Sep}.  Yet, it remains challenging to combine the accuracy of the first type of approaches with the computational cost of the second. For that purpose, machine-learning methods were proposed during the past couple of decades~\cite{Behler2007Apr,Bartok2010Apr,behler2015,Manzhos2020} and were successfully employed for various types of materials including pure metals~\cite{Novoselov2019Jun,Seko2015Aug,Takahashi2018Jun,Zeni2018Jun,Botu2017Jan}, oxides~\cite{Artrith2011Apr,Quaranta2017Apr}, water~\cite{Nguyen2018Jun,Bartok2013Aug,Morawietz2012Feb,Natarajan2016Oct,Morawietz2013Aug}, amorphous materials~\cite{Deringer2017Mar,Bartok2018Dec,Caro2018Apr,Deringer2018Jun,Deringer2018Nov,Sosso2018Jul} and hybrid perovskites~\cite{Jinnouchi2019Jun}. Similar strategies were also carried out for organic molecules~\cite{Bereau2018Jun,Sauceda2019Mar,Bartok2017Dec,Veit2019Apr,Smith2017Mar,Schutt2019Nov,Dral2020May,Eckhoff2019Jun}.

Recently, the machine-learning techniques have been applied in the context of the simulation of infrared spectra~\cite{Gastegger2017Sep,Lam2020Mar,kovacs2020}. For PAHs molecules, information regarding the vibrational frequencies are crucial in order to use optical spectroscopy to detect those molecules both in the interstellar medium~\cite{Sandford2013Mar,Piest1999Jun} and in flames~\cite{Ciajolo1998Jan}. Considering the specific inherent structures present in PAHs molecules, machine-learning approaches also raise the question of transferability from small to larger molecules. 

In this work, machine-learning techniques were employed to obtain: (a) a neural network potential energy surface and (b) a dipole mapping also based on a neural network architecture. Altogether, it enables us to compute infrared spectra of PAHs molecules, including anharmonic effects. In particular, we first built an accurate DFT database (energies and forces) for 11 neutral PAHs molecules which was then used to train both neural networks. From the learned potential energy surface and dipole mapping, we deduced the harmonic and anharmonic vibrational frequencies for 17 PAHs molecules. In this paper, we will first describe our methodology in particular the ANN parameters and the benchmarking molecules. Then, we will evaluate our approach by comparing the obtained results to fully quantum calculations and to experimental measurements.


\section{Methodology}
\label{method}
To build the ANN potentials and ANN dipole functions, data sets of reference energies, forces and dipoles for several PAHs molecules were generated using a DFT quantum chemical method. This same level of theory has been employed in the computation of the corresponding harmonic and anharmonic frequencies. The latter are required for the prediction of the IR spectra. In the next subsection, we will describe how such database is built and the structure of the employed ANN.
 
\subsection{Database}

The training of the ANN requires the generation of an extensive database, which in our case is constituted by equilibrium geometries of the PAH molecules together with their respective out-of-equilibrium derivatives. 
With respect to the former, geometries of molecules are optimized and IR frequencies are computed. 
Regarding the latter, each atom is displaced from its equilibrium position in the three Cartesian coordinates directions by $\pm$ 0.01, $\pm$ 0.02, $\pm$ 0.2, $\pm$ 0.3, and $\pm$ 0.5~\AA.
Then, single-point calculations are performed in order to obtain (i) the forces and energies, and (ii) the charges and dipoles for each of the optimized geometries as well as their deformed structures.
All quantum chemical calculations have been carried out using the N07D basis set\cite{Barone2008} and the B3LYP functional\cite{Stephens1994} as implemented in the Gaussian 09 \cite{G09} package.

In this work, we studied 17 molecules from the PAH family. 
Eleven of them have been integrated in the database for the training set of the considered ANN (Anthracene, Benzofluorene, Chrysene, Corannulene, Coronene, Fluorene, Naphthalene, Perylene, Phenalene, Phenanthrene, Pyrene),while the remaining six have been used as validation set (Benzoperylene, Benzopyrene, Ovalene, Pentacene, Tetracene, Triphenylene).
In Figure S1 of the Supporting Material, we present their structures. 

\subsection{The Artificial Neural Network}
In this work, we have used the neural network potential package, n2p2~\cite{Singraber2019Jan,Singraber2019Apr}. Such ANN is based on the method by Behler and Parrinello~\cite{Behler2007Apr} in which $N$ atoms are assumed to contribute separately to the total potential energy: $V = \sum_{j=1}^{N} E_j(G_j) $. 
The ANN structure is composed of three parts composed of multiple layers, which are used to calculate the contribution of the $j$-th atom to the total energy: 

(1) starting from the real atomic positions, symmetry functions $G_j$ are built to describe the local environment of each atom $j$. Eq.~\ref{G1} and~\ref{G2} refer, respectively, to the radial and the narrow angular symmetry functions constructed as a sum of Gaussian functions~\cite{Singraber2019Jan}, which by construction vanish beyond a cutoff distance $r_c$.
For the radial part (Eq.~\ref{G1}), two types of symmetry functions are included. On the one hand, the $r_s$ term is taken into account to describe the shift of a first radial function, and on another hand, a second radial function is added and centered on the atom with $r_s = 0$.
$\lambda$, $\eta$, and $\zeta$ are adjustable free parameters for all possible two-atom combinations. 
In radial and angular functions, $f_c(r_{ij})$ is a cutoff function taken as a hyperbolic tangent function. 
The resulting parameters are given in table~\ref{table:RadialSymFunc} for radial functions and table~\ref{table:AngularSymFunc} for angular ones.
Altogether, these symmetry functions form the input layer of the ANN. 

\begin{align}
   & G_j^1 = \sum_{j \ne i}^{N} e^{-\eta(r_{ij} - r_s)^2}f_c(r_{ij}) \label{G1} \\
   & G_j^2 = 2^{1-\zeta}\sum_{j,k \ne j}^{N} (1 + \lambda \cos(\theta_{ijk}))^{\zeta} \times e^{-\eta(r_{ij}^2 + r_{ik}^2 + r_{jk}^2)}f_c(r_{ij})f_c(r_{ik})f_c(r_{jk})  \label{G2} \\
   & \text{with},~f_c(r_{ij}) = \left\{   \nonumber
    \begin{array}{ll}
       \tanh^3 \left(1 - \frac{r_{ij}}{r_c}\right), & for~r_{ij} \leq r_c,  \\
       0,  & for~r_{ij} > r_c.
    \end{array} \right. \\ 
    & and,~ cos(\theta_{ijk}) = \frac{\vec{r_{ij}}.\vec{r_{ik}}}{r_{ij}r_{ik}} \nonumber
\end{align}

(2) The symmetry functions $G_j$ are injected in multiple hidden layers of $n_l$ nodes each. 
In this work, only two hidden layers have been employed. 
The output result $y_i^k$ of the neuron $i$ of the hidden layer $k$ is related non-linearly to the previous output result $y_{m}^{l}$ of the neuron $m$ of the layer $l = k - 1$, as shown in Eq.~\ref{yik}.
All nodes are thus connected with each node of the previous layer, and their results are weighted with a current bias $w_{0i}^k$, and a weight $w_{mi}^{lk}$. 
The non-linearity of the relationship is ensured thanks to an activation function $f_a^k$, which is taken as a hyperbolic tangent in this work.

\begin{align}
y_i^k = f_a^k \left(w_{0i}^k + \sum_{m = 1}^{n_l} w_{mi}^{lk}y_{m}^{l} \right) \label{yik}
\end{align}

(3) The output layer collects the results of the last layer of nodes, $i.e.$ the atomic energy $E_j$, and they are linearly combined to compute the total potential energy.
Atomic forces can be derived from the atomic energy and symmetry functions: $F_i = - \sum_{j=1}^{N} \sum_{k=1}^{N_j}  \frac{\partial E_j}{\partial G_{j,k}}\nabla_i G_{j,k}$.

The training of such ANN consists on computing the best set of weights and biases by fitting on the forces and energies of the database geometries, using the Kalman filter optimization method~\cite{Singraber2019Apr}.
Initial weights and biases are taken randomly and are adjusted from a new set of coordinates at each iteration. 
The training efficiency is evaluated by calculating the root-mean-square deviation (RMSD) on forces and energy. 
While 80~\% of data sets are allocated to the training of the ANN, the remaining 20~\% is used for the testing.

As previously discussed, energies are for the calculations of the vibrational frequencies, and dipole values are necessary for the intensity of these frequencies. 
Therefore, we have integrated in the n2p2 package an extra ANN for the dipole calculations in an analogous manner as in the case of the potential~\cite{Gastegger2017Sep}. 
In this approach, the molecular dipole $\overrightarrow {\mu}$ is considered as a sum of such environment dependent atomic partial charges:
\begin{align}
\overrightarrow {\mu} = \sum_{i = 1}^{N} q_i \overrightarrow{r_i}
\label{mu}
\end{align}
where $q_i$ is the charge of atom $i$ modeled by the ANN, and $\overrightarrow{r_i}$
is the vector position of the atom number $i$ of the molecule composed of $N$ atoms. 
We train the elemental ANN to reproduce the global charge of the molecules and the molecular dipole moments by minimizing a cost function defined as:
\begin{align}
C_Q = \frac{1}{M}  \sum_{j = 1}^{M} (Q^{ref}_j - Q_j)^2 + \frac{1}{3M} \sum_{j = 1}^{M} \sum_{k = 1}^{3} (\mu^{ref}_{jk}-\mu_{jk})^2
\end{align}

where $M$ is the number of molecules in the training set, $Q^{ref}_j$ and $\mu^{ref}_{jk}$ are the reference total charge and dipole moment components of the molecule $j$. 
$Q_j$ is the total charge obtained from ANN by $Q_j = \sum_{i = 1}^{N} q_i$, and $\mu_{jk}$ is the dipole moment given by equation \ref{mu}. 
The same symmetry functions are used in both the ANN energy potential and dipole.

\begin{table}
\caption{Radial Symmetry Function Parameters for $\eta/Bohr^{-2}$, $r_s$, and $r_c/Bohr$
}
\label{table:RadialSymFunc}
\begin{center}
\begin{tabular}{|l c  c|}
\hline
$\eta$ & $r_s$ & $r_c$ \\
\hline
    2.000000 &     0.000 &         12.0 \\
    0.116500 &     0.000 &         12.0 \\
    0.037680 &     0.000 &         12.0 \\
    0.018390 &     0.000 &         12.0 \\
    0.010860 &     0.000 &         12.0 \\
    0.007159 &     0.000 &         12.0 \\
    0.005072 &     0.000 &         12.0 \\
    0.003781 &     0.000 &         12.0 \\
  \hline
    0.202500 &     0.500 &         12.0 \\
    0.202500 &     2.071 &         12.0 \\
    0.202500 &     3.643 &         12.0 \\
    0.202500 &     5.214 &         12.0 \\
    0.202500 &     6.786 &         12.0 \\
    0.202500 &     8.357 &         12.0 \\
    0.202500 &     9.929 &         12.0 \\
    0.202500 &    11.500 &         12.0 \\
\hline
\end{tabular}
\end{center}
\end{table}

\begin{table}
\caption{Angular Symmetry Function Parameters for $\eta/Bohr^{-2}$, $\lambda$, $\xi$ and $r_c/Bohr$}
\footnote{We used the same angular parameters for all kind of atom types}
\label{table:AngularSymFunc}

\small{
\begin{center}
\begin{tabular}{|l c c c|}
\hline
$\eta$ & $\lambda$ & $\xi$ & $r_c$ \\
\hline
0.001 &	-1 & 4.0 & 12.0\\
0.001 &	+1 & 4.0 & 12.0\\
0.010 &	-1 & 4.0 & 12.0\\
0.010 &	+1 & 4.0 & 12.0\\
0.030 &	-1 & 1.0 & 12.0\\
0.030 &	+1 & 1.0 & 12.0\\
0.070 &	-1 & 1.0 & 12.0\\
0.070 &	+1 & 1.0 & 12.0\\
\hline
\end{tabular}
\end{center}
}
\end{table}

\subsection{Calculation of infrared frequencies}
Harmonic frequencies and its anharmonic corrections have been computed within the explicit framework of generalized second-order vibrational perturbation theory (GVPT2)~\cite{Barone2005Jan,Barone2014Jan}, as implemented in our own code called iGVPT2~\cite{Barnes2016} that was interfaced with n2p2. Altogether, this allows us to compute the IR spectrum with the ANN potential energy and dipole. 

In particular, after optimization of the geometry, the harmonic frequencies and the normal modes are computed. For that purpose, cubic and quartic derivatives of energy are determined numerically using Yagi et al. method\cite{yagi00}. Then we approximate the PES as a quartic force field (QFF) :

\begin{align}
& V(\textbf{Q}) = V_0 + V_1(\textbf{Q}) + V_2(\textbf{Q}) + V_3(\textbf{Q})  \\
& V_1(\textbf{Q}) =  \sum_{i=1}^{f} \frac{1}{2}h_i Q_i^2 + \frac{1}{6}t_{iii} Q_i^3   + \frac{1}{24}u_{iiii} Q_i^4  \\
& V_2(\textbf{Q}) =  \sum_{ij, i\neq j}^{f} \frac{1}{2}t_{ijj}Q_iQ_j^2 + \frac{1}{6}u_{ijjj} Q_iQ_j^3   + \sum_{ij, i<j}^{f} \frac{1}{4}u_{iijj} Q_i^2Q_j^2  \\                      
& V_3(\textbf{Q}) = \sum_{ijk, i\neq j<k}^{f} t_{ijk}Q_iQ_jQ_k + \sum_{ij, i \neq j<k}^{f} \frac{1}{2}u_{iijk} Q_i^2Q_jQ_k
\end{align}
where $h_i$, $t_{ijk}$ and $u_{ijkl}$ correspond to the second-, third-, and fourth-order derivatives of the energy $V_0$, expressed in terms of normal coordinates \textbf{Q} associated to the $f = 3N - 6$ ($3N - 5$ for a linear molecule) normal modes, with $N$ being the number of atoms. a normal mode.
Calculations of the second- and third-order derivatives of the dipole are also performed to obtain the IR intensities.


\section{Results}
\label{results}

Three ANN architectures composed of two hidden layers have been trained. For each we considered different number of nodes per layer, $i.e.$ 15, 20, and 30 nodes per layer.
The training of the potential and dipole ANNs results on a significant convergence characterized by RMSDs shown in table~\ref{tab:rmsd}. 
The main feature of these ANNs is that the fit accuracy is below 0.7 meV for energy and below 60 meV/\AA~for the forces, both for training and validation sets. 
When increasing the number of nodes per layer, the RMSD decreases by a maximum of $\sim$~40~\% for energy, and $\sim$~24~\% for forces.
RMSD on charges decreases by $\sim$~10~\% with the number of nodes per layer, while the RMSD on dipole increases slightly. 

In the following, we will present and compare harmonic and anharmonic frequencies obtained from our ANN together with the resulting IR spectra.

\begin{table}
\centering
\caption{RMSD obtained for the energies, the forces, the charges, and the molecular dipoles when training the ANN potential and the ANN dipole with different number of nodes on two layers.}
\label{tab:rmsd}    
\begin{tabular}{| p{6em} p{6em} p{6em} p{6em} p{6em} |}
\hline
Nodes & Energy (meV) & Forces (meV/\AA)  & Charges (a.u.) ($\times 10^{-4})$ & Dipole (D) ($\times 10^{-3})$\\
\hline
15 $\times$ 15 & 0.69 & 59.82 & 5.72 & 3.22 \\
20 $\times$ 20 & 0.41 & 51.04 & 5.20 & 3.29 \\
30 $\times$ 30 & 0.55 & 45.49 & 5.18 & 3.74 \\
\hline
\end{tabular}
\end{table}

\subsection{Harmonic frequencies}
\label{results:1}

Using the resulting ANN potential, harmonic frequencies for each normal mode have been computed and compared with those calculated using DFT. 
Low ($\leq$ 2000 cm$^{-1}$) and high ($>$ 2000 cm$^{-1}$) frequency ranges are differentiated from the whole range of frequencies. 
Statistical errors, namely RMSD, mean absolute error (MAE), averaged sign error (ASE), and frequency maximum (UMAX), are estimated by considering all the dataset, for the three frequency ranges, and for each ANN (see table~\ref{tab:stats_harm}). 

\begin{figure*}[ht]
    \centering
    \includegraphics[width=8.6cm]{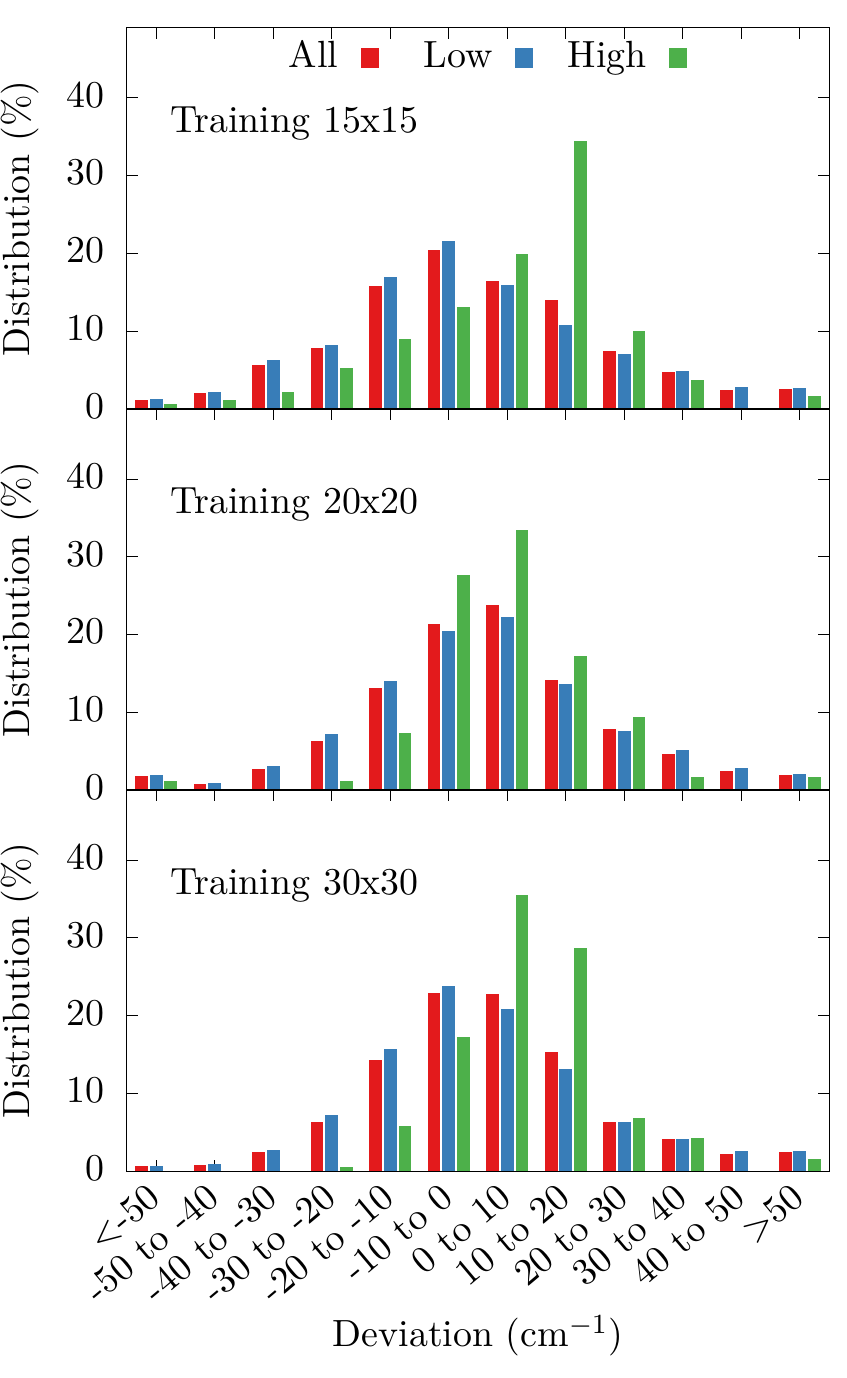}
    \caption{Distribution of the deviations between the harmonic frequencies computed with the ANN and the DFT calculations. Frequencies are obtained from the ANN trained with (top) 15 $\times$ 15 nodes, (middle) 20 $\times$ 20 nodes, and (bottom) 30 $\times$ 30 nodes.}
    \label{fig:AllHarm}
\end{figure*}

To further compare both sets of data, distributions of the deviation between ANN and DFT results are represented in the form of histograms in Figure~\ref{fig:AllHarm} for the three trained ANNs.
As it can be observed the distributions are more peaked when increasing the number of nodes per layer. 
Indeed, from 15 to 30 nodes per layer, deviations between -10 and 10 cm$^{-1}$ passing from 50.8 to 60.8~\%.
However, almost the entire distribution of the frequencies, $i.e.$ 86-92~\%, are predicted between -30 and 30 cm$^{-1}$, for all the trained ANN. 
A common characteristic is that the distributions of the high frequencies are greatly peaked and shifted towards positive deviations with ASEs between 4.8 and 8.6 cm$^{-1}$, whereas the low frequencies follow the trend of the whole range of frequencies, centered around 0. 
This indicates that the ANN potential slightly underestimates high frequencies compared to those calculated using DFT.

\begin{figure*}[htb!]
    \centering
    \includegraphics[width=16cm]{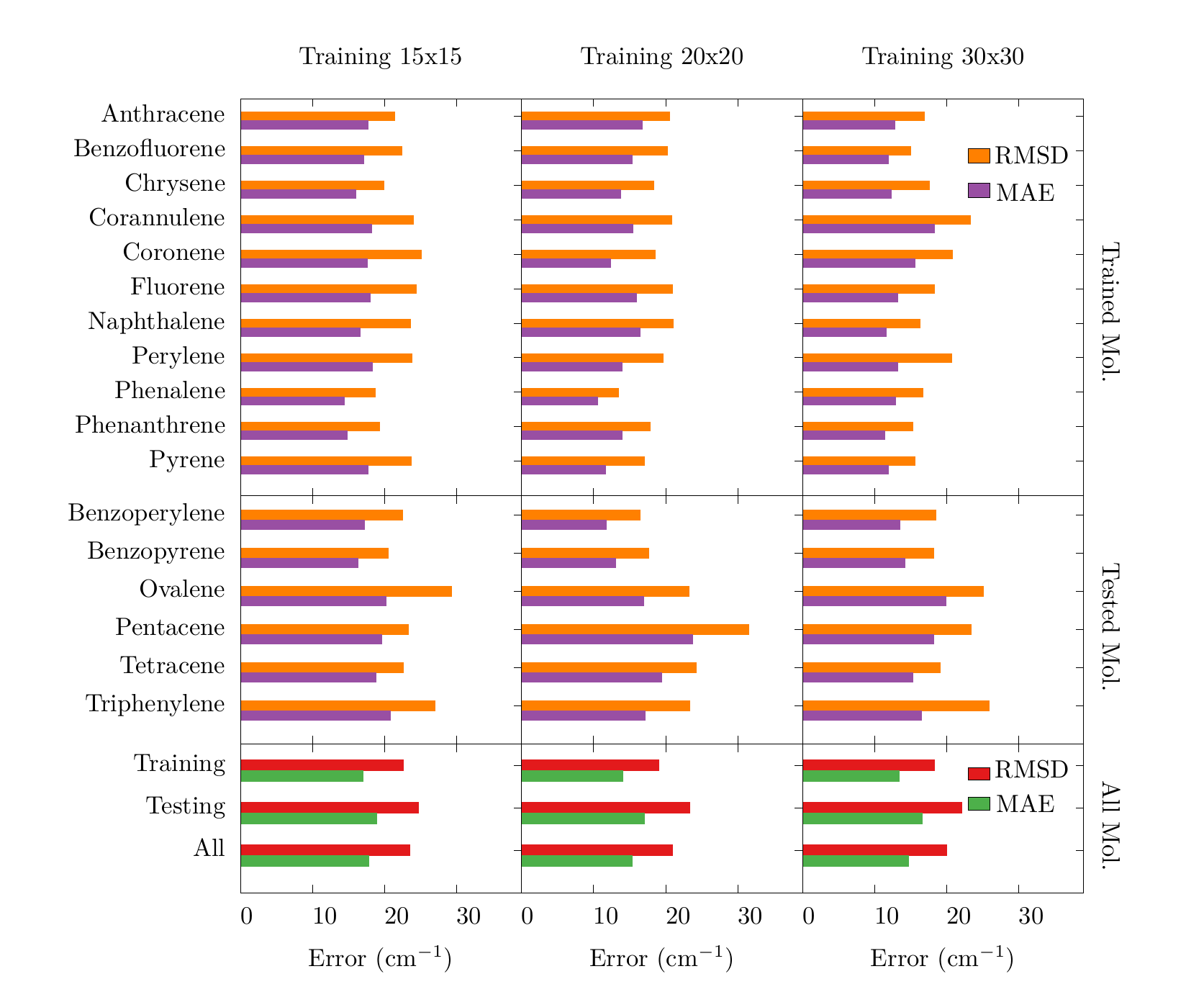}
    \caption{Root-mean-square deviations (RMSD) and Mean Absolute Errors (MAE) of harmonic vibrational frequencies obtained by the full ANN calculations relative to the full DFT calculations for each molecule. Results obtained with the three trained ANN structures are presented.}
    \label{fig:MolHarm}
\end{figure*}

At the individual scale, RMSDs and MAEs for all the frequencies are displayed for each molecule and for each ANN system in figure~\ref{fig:MolHarm}. 
Both trained and tested structures present frequency errors oscillating between 19 and 24.7 cm$^{-1}$, for all ANNs.
Increasing the number of nodes per layer slightly increases the accuracy of the predicted frequencies. 
A small decrease of the RMSDs for all the frequencies exists from 23.5 to 21.0 and 20.0 cm$^{-1}$ for the ANNs of 15, 20 and 30 nodes per layer, respectively.
Similarly, the RMSDs of the trained molecules are improved, passing from 22.6 to 19.0 and 18.3 cm$^{-1}$ with the increase of nodes per layer.
For the tested molecules, their RMSDs are less reduced, $i.e.$ 24.7, 23.4, and 22.1 cm$^{-1}$ for the ANNs composed of 15, 20 and 30 nodes per layer, respectively.
The molecules with the higher RMSDs are generally the ovalene, the pentacene, and the triphenylene independently from the considered ANN system. 
These molecules have large structures, and have not been included in the training database. 

In general, we observe a slight improvement in the quality of the calculations of harmonic frequencies when increasing the number of nodes per layer, even though improvements are more important from 15 to 20 nodes per layer, instead of from 20 to 30 nodes per layer.
Moreover, low frequencies are better predicted than their high counterparts, since high frequencies are more difficult to predict. 
For an overall outlook, spectra of harmonic frequencies calculated with the 30 $\times$ 30  ANN and using DFT are displayed in figures S2 and S3 in the Supp. Mat.

\begin{table}
\centering
\caption{Statistical Errors (in cm$^{-1}$ ) using all harmonic frequencies, low frequencies ($\leq$ 2000 cm$^{-1}$ ), and high frequencies ($>$ 2000 cm$^{-1}$), for all the trained ANN.}
\label{tab:stats_harm}    
\begin{tabular}{| p{6em} p{6em} p{6em} p{6em} p{6em} p{6em} |}
\hline
ANN & Frequencies & RMSD  & MAE  & ASE & UMAX \\
\hline
\multirow{3}{*}{15 $\times$ 15} & Low & 23.6 & 18.1 & -0.9 & 93.7 \\
                                & High & 22.7 & 15.9 & 7.0 & 139.7 \\
                                & All & 23.5 & 17.8 & 0.2 & 139.7 \\

    &  &  &  &  &  \\
\multirow{3}{*}{20 $\times$ 20} & Low & 21.7 & 16.1 & 1.8 & 102.5 \\
& High & 15.8 & 10.7 & 4.8 & 81.2 \\
& All & 21.0 & 15.3 & 2.2 & 102.5 \\

    &  &  &  &  &  \\
\multirow{3}{*}{30 $\times$ 30} & Low & 20.6 & 15.2 & 1.6 & 95.3 \\
& High & 15.3 & 11.7 & 8.6 & 58.9 \\
& All & 20.0 & 14.7 & 2.5 & 95.3 \\

\hline
\end{tabular}
\end{table}

\subsection{Fundamental frequencies}
\label{results:2} 

Fundamental frequencies have been calculated using the GVPT2 approach which includes anharmonic effects in harmonic frequencies previously calculated. 
Similarly, fundamental frequencies calculated using the ANN systems are compared with those calculated using DFT. 
Please note that due to technical problems in DFT calculations and due to the consequent computational time, we were not able to calculate correctly the fundamental frequencies of the coronene and ovalene molecules, and we did not take into account them in the following results.

\begin{figure*}[ht]
    \centering
    \includegraphics[width=8.6cm]{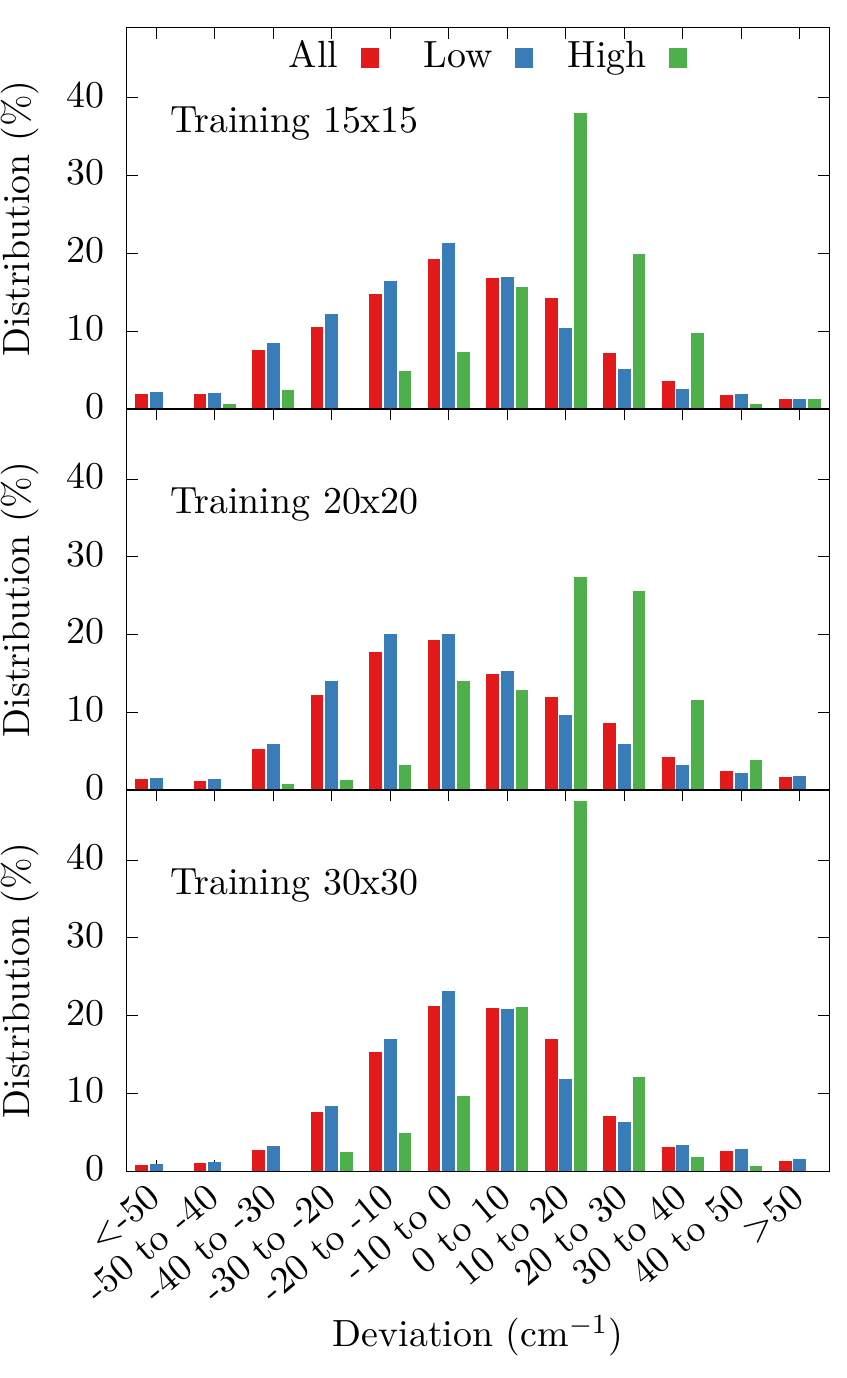}
    \caption{Distribution of the deviations between the vibrational frequencies computed with the ANN and the DFT calculations. Frequencies are obtained from the ANN trained with (top) 15 $\times$ 15 nodes, (middle) 20 $\times$ 20 nodes, and (bottom) 30 $\times$ 30 nodes.}
    \label{fig:AllAnh}
\end{figure*}

Again for the fundamental frequencies, displayed in figure~\ref{fig:AllAnh}, we find a similar distribution as in the case of the harmonic frequencies.
High frequencies are even more peaked and shifted towards the positive frequencies with ASEs of 13.6, 15.4, and 10.5 cm$^{-1}$ for the 15, 20, and 30 nodes-per-layer ANN system (see table~\ref{tab:stats_anh}). 
Low frequencies are slightly overestimated compared to the DFT ones, $i.e.$ negative ASEs of -5.6 and -4.3 cm$^{-1}$ for the 15 $\times$ 15 and 20 $\times$ 20 ANN systems.
Only the low frequencies of the 30 $\times$ 30 ANN are symmetrically distributed and centered around 0, even when taking the whole ranges of frequencies for all the ANN architectures.
Surprisingly, RMSDs for all the fundamental frequencies are as low as those calculated for the harmonic frequencies, except for the ANN system with 20 nodes per layer. 
Indeed, RMSDs of 22.7, 22.0 and 19.8 cm$^{-1}$ are obtained when increasing the nodes per layer from 15 to 20, and 30, respectively (table~\ref{tab:stats_anh}).

\begin{figure*}[ht]
    \centering
    \includegraphics[width=16cm]{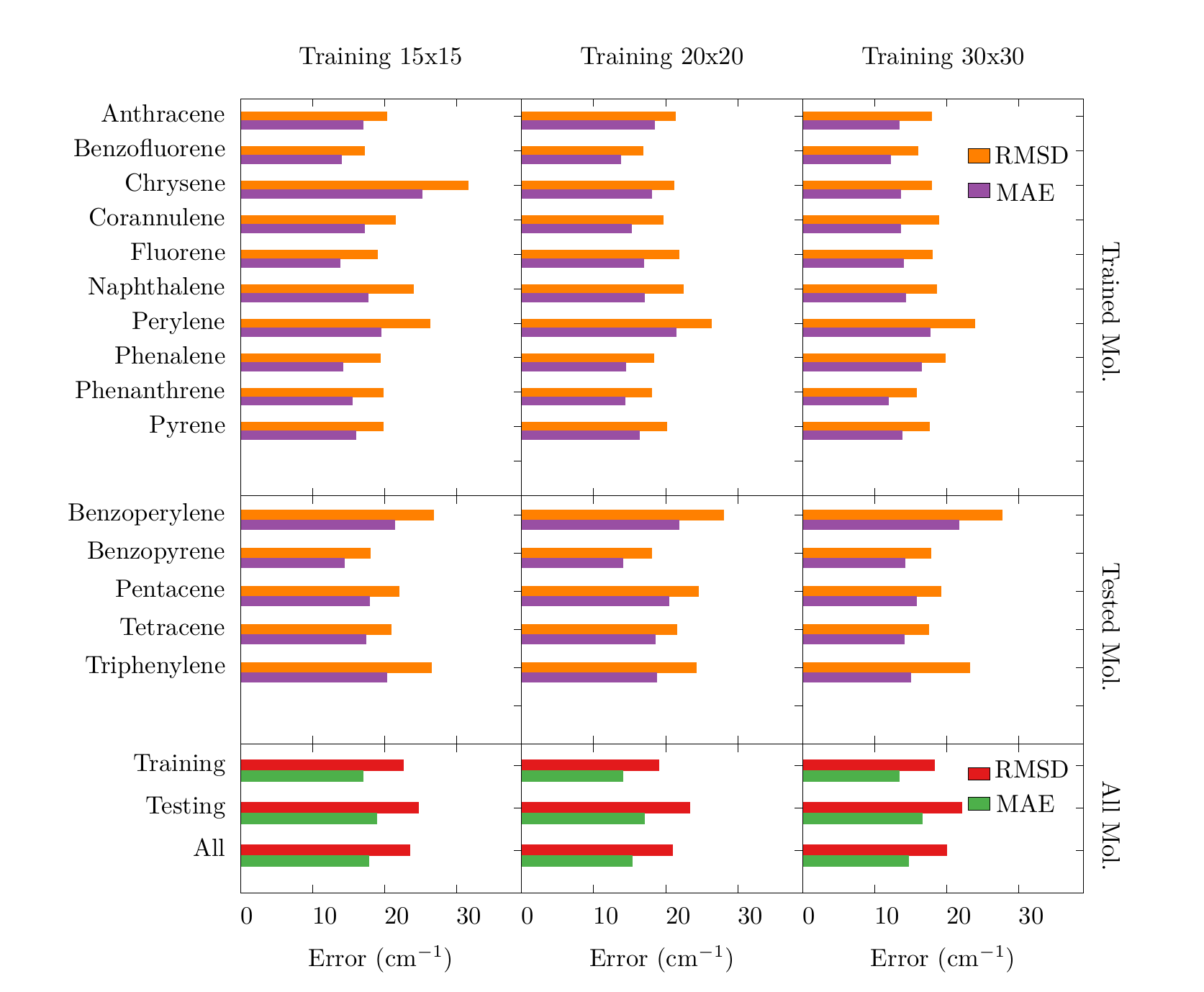}
    \caption{Root-mean-square deviations (RMSD) and Mean Absolute Errors (MAE) of fundamental vibrational frequencies obtained by the full ANN calculations relative to the full DFT calculations for each molecule. Results obtained with the three trained ANN structures are presented.}
    \label{fig:MolAnh}
\end{figure*}

In figure~\ref{fig:MolAnh}, we can see clearly that increasing the nodes per layer leads to the reduction of the RMSDs, especially for the trained molecules, where RMSDs go from 22.6 to 19.1 and 18.3 cm$^{-1}$ when using respectively 15, 20 and 30 nodes per layer.  
These RMSD reductions are more visible for some molecules such as the anthracene, the chrysene or the phenanthrene molecules. 
However, anharmonic frequencies of some molecules could not be improved with such strategy, $e.g.$ the perylene or the benzoperylene molecules. 
Such molecules have special hexagonal aromatic ring found also in coronene and ovalene which make the DFT calculations difficult.

Nevertheless, using ANN architecture enables to calculate efficiently fundamental frequencies for trained molecules, and to transfer it to the set of untrained molecules.
Using an ANN with 30 nodes per layer seems to be a better compromise this time, whereas 20 nodes per layer seemed to be enough when calculating harmonic frequencies.

\begin{table}
\centering
\caption{Statistical Errors (in cm$^{-1}$ ) using all fundamental frequencies, low frequencies ($\leq$ 2000 cm$^{-1}$ ), and high frequencies ($>$ 2000 cm$^{-1}$), for all the trained ANN.}
\label{tab:stats_anh}    
\begin{tabular}{| p{6em} p{6em} p{6em} p{6em} p{6em} p{6em} |}
\hline
ANN & Frequencies & RMSD  & MAE  & ASE & UMAX \\
\hline
\multirow{3}{*}{15 $\times$ 15} & Low & 23.0 & 17.6 & -5.6 & 93.9 \\
& High & 21.0 & 17.8 & 13.6 & 70.7 \\
& All & 22.7 & 17.6 & -2.9 & 93.9 \\

    &  &  &  &  &  \\
\multirow{3}{*}{20 $\times$ 20} & Low & 22.1 & 17.4 & -4.3 & 82.8 \\
& High & 21.5 & 18.5 & 15.4 & 44.7 \\
& All & 22.0 & 17.5 & -1.6 & 82.8 \\

    &  &  &  &  &  \\
\multirow{3}{*}{30 $\times$ 30} & Low & 20.4 & 15.2 & 0.0 & 93.3 \\
& High & 15.6 & 13.6 & 10.5 & 48.0 \\
& All & 19.8 & 15.0 & 1.5 & 93.3 \\

\hline
\end{tabular}
\end{table}

\subsection{Spectra}
\label{results:3}

\begin{figure*}[ht]
    \centering
    \includegraphics[width=8.6cm]{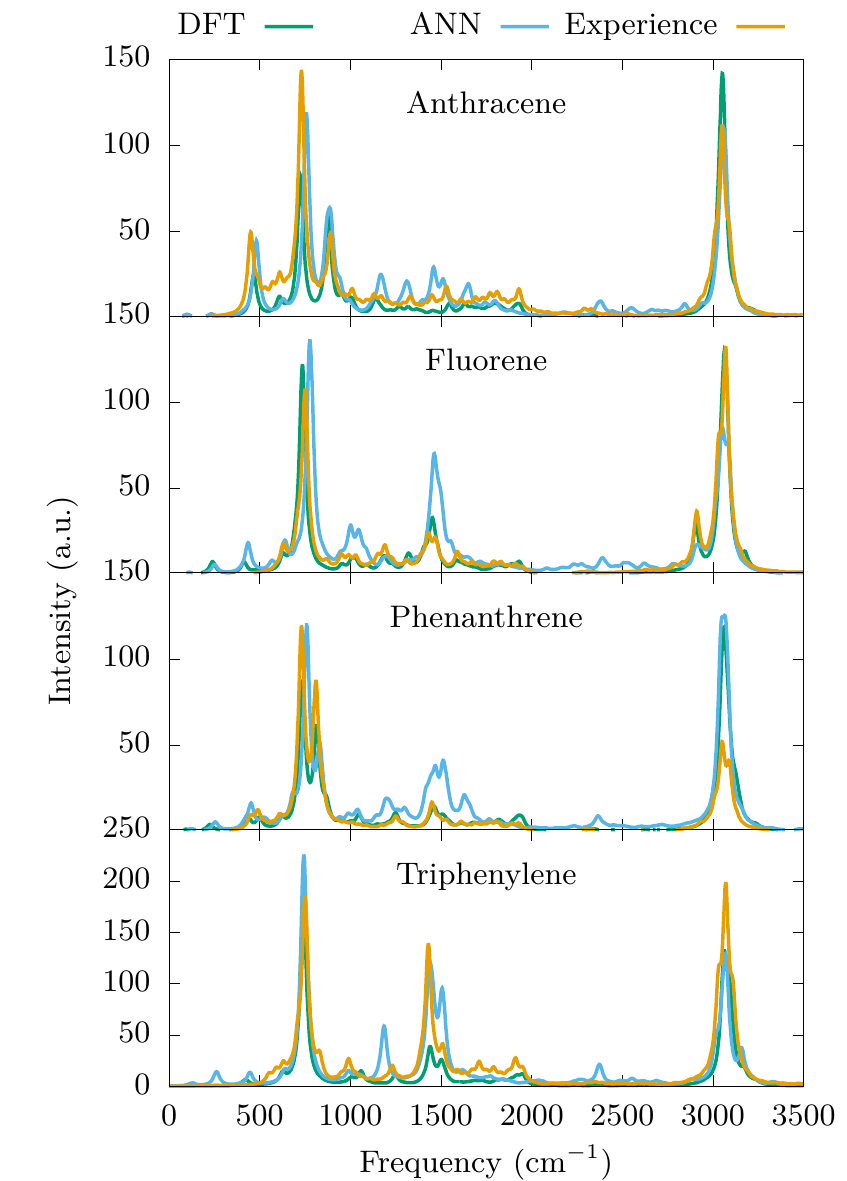}
    \caption{IR spectra of anthracene, fluorene, phenanthrene, and triphenylene molecules calculated using the 30 $\times$ 30 ANN system, and compared with experimental spectra and those calculated using DFT.}
    \label{fig:spectra}
\end{figure*}

In the previous section, only fundamental frequencies were considered and compared with DFT results. 
To go further, we take now into account all the frequencies including overtones and combination bands of the anharmonic frequencies. 
With this information, we can proceed to plot the IR spectrum for each molecule.
IR spectra of anthracene, fluorene, phenanthrene, and triphenylene molecules are calculated using the 30 $\times$ 30 ANN system and displayed in figure~\ref{fig:spectra}. 
For comparison, we added spectra obtained from our DFT calculations as well as experimental data taken from the National Institute of Standards and Technology database website~\cite{NIST} (when available). 
To better compare with experimental spectra, the same order of magnitude is reached by reducing experimental intensities by the ratio $I_{max}^{EXP}/I_{max}^{DFT}$, with $I_{max}^{EXP}$ and $I_{max}^{DFT}$ the maximum of intensity of experimental data and DFT calculations, respectively.
IR spectra of all the molecules calculated with the 30 $\times$ 30 ANN are presented in Supp. Mat. in figures S4 and S5. 

In overall, IR spectra calculated using ANN are in excellent agreement with DFT and experimental spectra. 
Main peaks are well represented both in frequency and in intensity - when compared only with DFT spectra - with slightly shifts for peaks around 750 cm$^{-1}$ for the anthracene and the fluorene molecules. 
Such strong peak corresponds to the out-of-plane C-H bending mode, clearly predominant in such molecules. 
Even in the case of the phenanthrene molecule spectrum, this mode is decoupled in two peaks at 745 and 826 cm$^{-1}$ and they are well predicted by the ANN. 
One tendency of the ANN system is to overestimate some satellite peaks below 500 cm$^{-1}$ or between 1000 and 2000 cm$^{-1}$.
For instance, peaks at 1500 cm$^{-1}$ coincide with in-plane C-H bending mode.


\subsection{Computational time}
\label{results:4}

\begin{figure*}[ht]
    \centering
    \includegraphics[width=8.6cm]{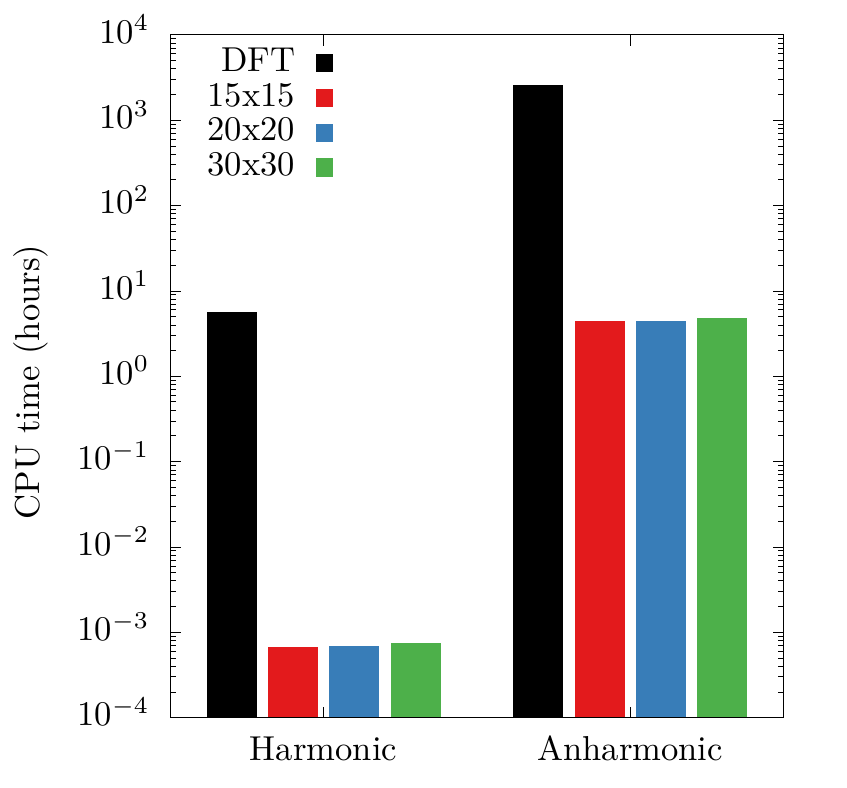}
    \caption{Comparison of computational times averaged over all the molecules when using DFT or ANN architectures for harmonic and anharmonic frequencies calculations. }
    \label{fig:time}
\end{figure*}

As final remark, it should be highlighted that in addition to approach the accuracy of the DFT calculations, the computational time is also largely improved.
By averaging the CPU times for all the molecules (Fig.~\ref{fig:time}), a gain of 3 orders of magnitude is obtained when calculating anharmonic frequencies against DFT computational time, and reached 4 orders of magnitude for the harmonic frequencies calculations, independently of the number of nodes per layer. 
For instance, by taking one of our largest molecules, namely the corannulene molecule, while the calculation of the anharmonic frequencies using DFT lasts more than 142 days of CPU time, the same calculation using our trained ANN systems lasts in average 3h40mns. 
Please note that CPU time only includes the time needed to calculate IR frequencies, while the time of the training is not taken into account. 


\section{Conclusions}
In summary, we developed artificial neural network systems to evaluate the IR spectra of PAH molecules. The obtained ANN systems are able to predict IR spectra combining DFT accuracy and low computational cost, from small to large PAH molecules. Indeed, three different ANNs composed of two hidden layers of 15, 20, and 30 neurons per layer were trained by using a database including 8 863 energies and 735 447 forces of 11 PAH molecules. Then, the computed vibrational frequencies of 19 PAH molecules are compared with DFT calculations. In overall, the obtained ANN architectures lead to excellent agreement with IR spectra both simulated by DFT and obtained in experiments. The obtained frequency RMSD and MAE are respectively equal to 19.8 and 15 cm$^{-1}$, when including all the molecules for the 30 $\times$ 30 ANN system. Increasing the number of nodes per layer from 15 to 20 and 30 shows a decrease of the frequency RMSD from 22.7 to 22 and 19.8 cm$^{-1}$, respectively. Moreover, IR spectra of molecules not included in the training are successfully reproduced, including more exotic PAH molecules in their structure such as the benzoperylene or the triphenylene. Such results are promising for the development of ANNs based on simple structures to extrapolate towards larger and more complex ones. In addition to approaching the accuracy of the DFT calculations, the computational cost is also largely improved. ANN CPU time is three orders of magnitude lower than DFT CPU time which allows for computing IR spectra of large and complex PAH molecules in few hours, compared to several days when using DFT methods. Finally, despite the overall good prediction of the ANN IR spectra, molecules with special bonds, such as the corannulene or the benzofluorene with their pentagonal aromatic cycle are not as well described by our modelling which paves the way for further improvements. 


\begin{acknowledgements}
\label{Acknowledgments}

This work was granted access to the HPC resources of the the "Centre de calcul CC-IN2P3" at Villeurbanne, France.
DP gratefully acknowledges computational support Andrei Borissov (ISMO, Paris-Saclay). JL acknowledges financial support of the Fonds de la Recherche Scientifique - FNRS. 
 
\end{acknowledgements}

%
 \section*{Conflict of interest}
 The authors declare that they have no conflict of interest.


%
%

\clearpage

\bibliographystyle{spphys}
\bibliography{biblio}


\end{document}